\documentclass[a4paper,11pt]{article}
\usepackage{jinstpub} 
\usepackage[percent]{overpic}
\usepackage{siunitx}

\title{\boldmath Physical Thickness Characterization of the FRIB Production Targets}

\author[a]{D. J. Lee}
\author[a]{M. Reaume}
\author[a]{W. Franklin} 
\author[a,1]{and J. Song\note{Corresponding author.}}
\affiliation[a]{Facility for Rare Isotope Beams, East Lansing, MI 48824, USA}       
\emailAdd{songj@frib.msu.edu}

\abstract{
  The FRIB heavy-ion accelerator, commissioned in 2022, is a leading facility
  for producing rare isotope beams (RIBs) and exploring nuclei beyond the limits of stability.
  These RIBs are produced via reactions between stable primary beams and a graphite target.
  Approximately 20–40$\%$ of the primary beam power is deposited in the target,
  requiring efficient thermal dissipation.
  Currently, FRIB operates with a primary beam power of up to 20 kW. To enhance thermal dissipation efficiency,
  a single-slice rotating graphite target with a diameter of approximately 30 cm is employed.
  The effective target region is a 1 cm-wide outer rim of the graphite disc.
  To achieve high RIB production rates, the areal thickness variation must be constrained within 2 $\%$.
  This paper presents physical thickness characterizations of FRIB production targets with various nominal thicknesses,
  measured using a custom-built non-contact thickness measurement apparatus.
}
\begin{document}
\maketitle
\flushbottom

\section{Introduction}\label{sec:intro}

\begin{figure}[h]
  \begin{center}
    \begin{overpic} [width=0.45\textwidth] {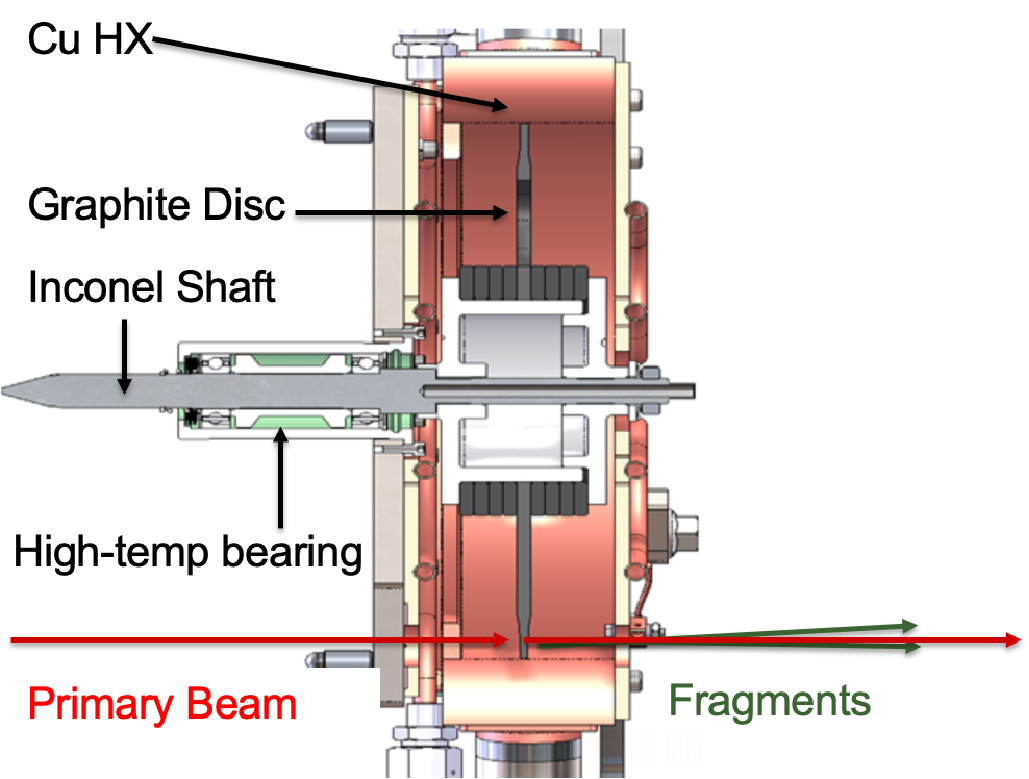}
    \end{overpic}
    \caption{
      Schematic layout of the FRIB single-slice rotating graphite target assembly, showing the graphite disk,
      Inconel drive shaft, high-temperature bearings, and the indirect water-cooled copper heat exchanger (HX).
    }
    \label{fig:0}
  \end{center}
\end{figure}

The Facility for Rare Isotope Beams (FRIB), a leading laboratory for rare isotope science, is undergoing a
staged ramp-up in primary beam power toward its design goal of 400 kW. Since beginning operations in 2022,
FRIB has delivered high-intensity stable ion
beams ranging from oxygen to uranium, progressively increasing its operational beam power from 1
kW to 20 kW \citep{int_1,int_2}, with current operations approaching 30 kW.
This capability supports a broad spectrum of nuclear science experiments,
encompassing investigations of nuclear structure, reaction mechanisms, and astrophysical phenomena.
To manage the substantial thermal loads generated during rare isotope production, FRIB employs a rotating
graphite target configured as a single-slice disc system \citep{int_3_1,int_3_2,int_3_3}.
The target system, shown in Figure \ref{fig:0}, consists of a 30 cm-diameter rotating graphite disc with graphite spacers,
supported by a YSZ ceramic hub, high-temperature bearings, and an Inconel drive shaft,
and enclosed by a copper heat exchanger.
The disc assembly provides a total of nine slots to enable future multi-slice disc operation at higher beam power.
At present, eight single-slice graphite discs with different thicknesses (1.2, 2.1, 3.5, 5, 8, 10, 15, and 18 mm)
are used for secondary beam production.
In the near future, sub-mm-thick targets are planned for operation at higher beam power.
The operational limit of the graphite disc is governed by sublimation, with a temperature limit of $\approx$ 1900 $^{o}$C
\cite{int_4}.
At a primary beam power of 20 kW and a rotational speed of 500 rpm,
the current maximum operating temperature is in the range of 800–1200 $^{o}$C,
depending on the target thickness.
Eight of these slots are filled with spacers.

High-power accelerator facilities use a variety of graphite grades for production targets,
including POCO ZXF-5Q \citep{int_5_1,int_5_2} at Fermilab,
Toyo Tanso IG-430 \citep{int_5_3,int_5_4} at J-PARC,
and SGL R6400 \citep{int_5_5,int_5_6} at PSI.
For FRIB, the graphite grade was selected based on several factors, including the minimum grain size (5 $\mu$m)
required to fabricate a sub-mm-thick disc with a diameter of 30 cm,
the availability of of sufficiently large blanks, machining feasibility, and relevant material properties
such as density (1.7-2.0 g/cm$^{3}$) and thermal conductivity ($>$80 W/mK). 
A finer-grained graphite, such as POCO ZXF-5Q with a grain size of 1 $\mu$m,
is not available in dimensions suitable for a 30-cm-diameter target.
In this work, Mersen graphite (grades 2160 and 2360) and POCO graphite (AXF-5Q) were used, all with a grain size of 5 $\mu$m.
The current FRIB target discs use Mersen grades 2160 and 2360.
POCO AXF-5Q was evaluated as a potential future option.
The inner surface of the heat exchanger is coated with
a high-emissivity layer ($\epsilon \approx$ 0.9) to enhance radiative cooling of the graphite disc.
High-temperature bearings and lubricants ($T_{max}$ = 280 $^{o}C$)
are used to accommodate power ramp-up and high-power operation.
Engineered to endure beam powers up to 50 kW,
the target mitigates thermal stress by rotating the disc, thereby spreading the energy deposition across
a wider surface area.
This approach lowers the surface power density and improves thermal efficiency. The current implementation
features a graphite disc approximately 30 cm in diameter, rotating at moderate speeds (500 rpm). Plans are
underway to increase the rotational speed to 2000 rpm in future operations, which will further suppress
the peak temperature and minimize temperature gradients within the irradiated region of the disc.

\begin{figure}
  \begin{center}
    \begin{overpic}[width=0.51\textwidth]{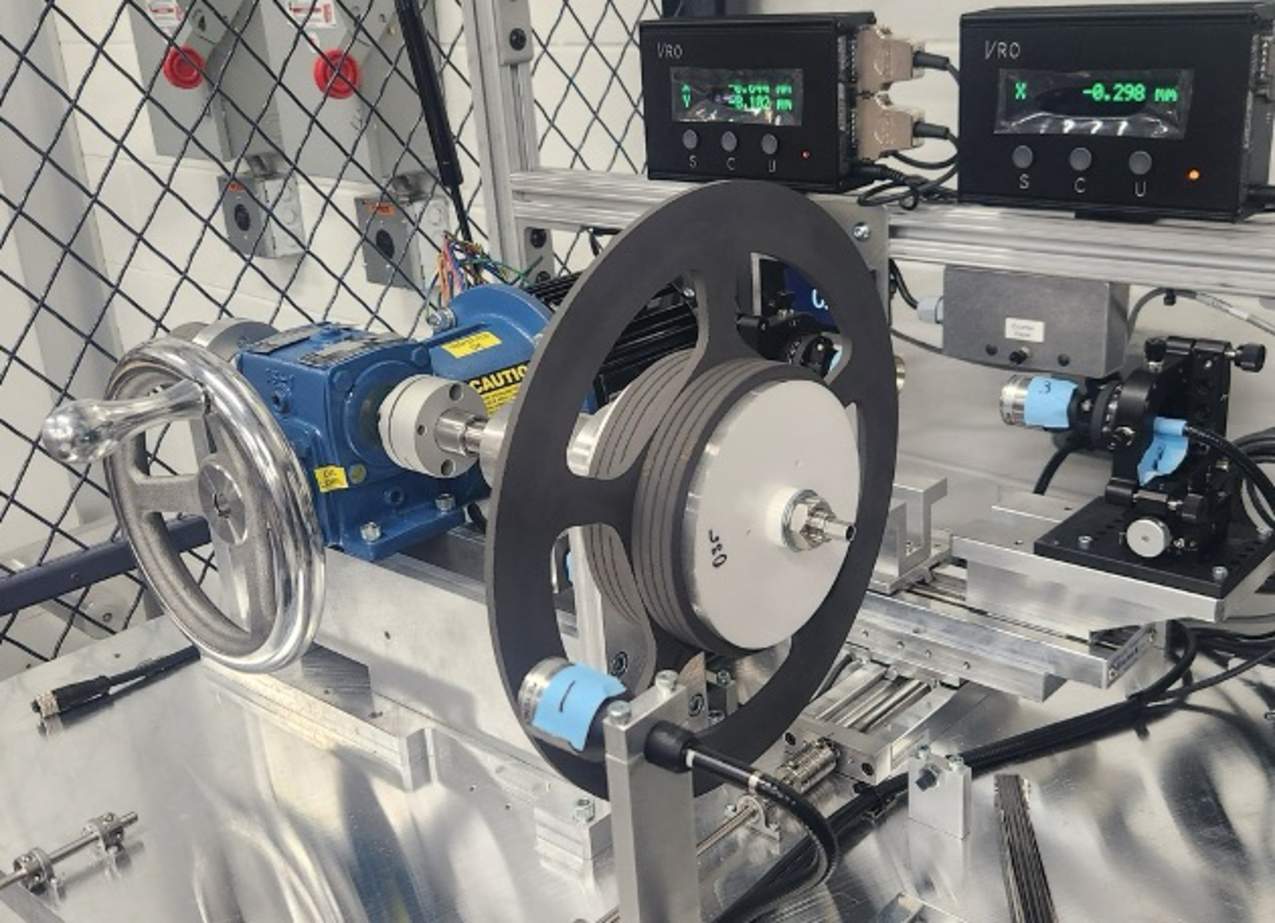}
    \end{overpic}
    \begin{overpic}[width=0.375\textwidth]{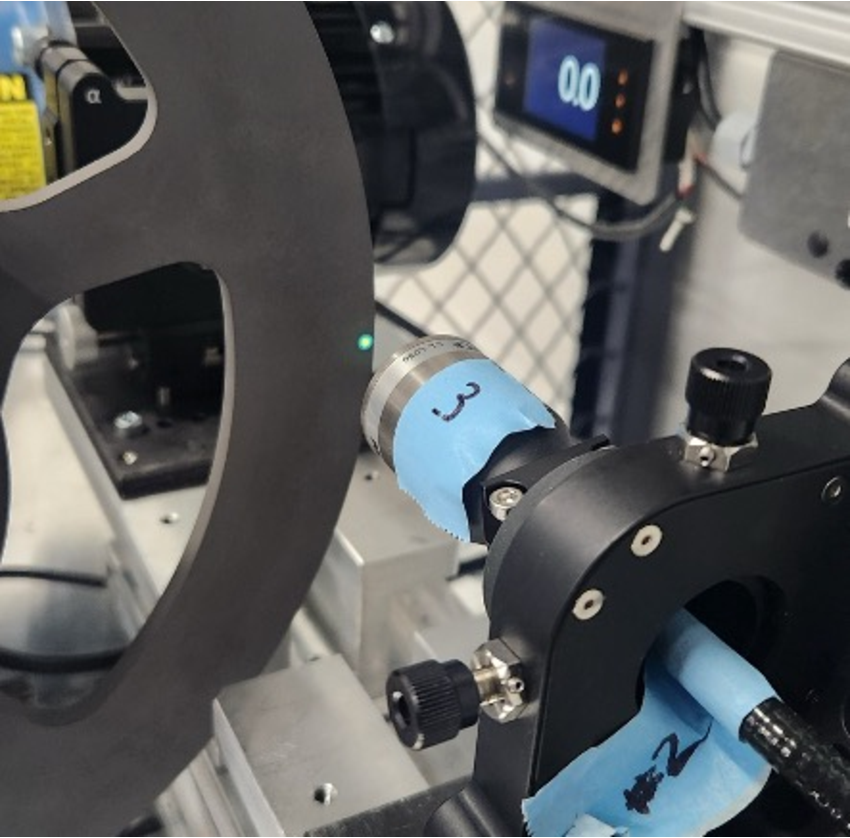}
      \put(-139,91){\colorbox{white}{\small a)}}
      \put(-125,18){\colorbox{white}{\small rotational unit}}
      \put(-95,85){\colorbox{white}{\small disc module assembly}}
      \put(-60,33){\colorbox{white}{\small displacement sensor}}
      \put(1,91){\colorbox{white}{\small b)}}
    \end{overpic}
    \caption{
      a) Photo of the test apparatus used for thickness measurement. The setup consists of a rotational unit,
      disc module assembly and displacement sensors. b) Detailed view showing the thickness measurement during
      target rotation. A green laser emitted from two opposing displacement sensor with a spot diameter of
      0.5 mm. The measured thickness is averaged over the illuminated area.
    }
    \label{fig:1}
  \end{center}
\end{figure}

The stable beam, accelerated in the FRIB accelerator section, is delivered to the target with
a beam spot size of 0.25 mm (1$\sigma$) in both horizontal and vertical directions. The beam irradiates
a 1 cm-wide circular region illustrated in Figure \ref{fig:0} near the rim of the rotating graphite disc.
The region maintains a uniform thickness, with the total allowable deviation limited to
less than 2$\%$.
This allowable thickness variation is constrained by the fixed momentum acceptance
of the Advanced Rare Isotope Separator (ARIS) \citep{int_6_1,int_6_2}.
To maintain uniform areal density in the irradiated region, local variations arising from material inhomogeneity,
porosity, or machining defects must be minimized, as they broaden the momentum distribution of the fragments
and reduce the fraction transmitted through ARIS to the experimental hall.
Based on measurements using actual beam irradiation, the combined uncertainty,
primarily due to porosity-related inhomogeneity and deformation during 500 rpm rotation,
is estimated to be within $\pm$ 1$\%$.
An additional contribution to the overall uncertainty
arises from local variations in physical thickness introduced during the machining process.
Currently, the graphite disc is replaced after every ten experimental runs.
To ensure consistent
experimental results, accurate characterization of the target thickness is essential. Previously,
thickness measurements were conducted at only five discrete positions using a micrometer, introducing
significant uncertainty and limited accuracy. To address this limitation, a continuous thickness measurement apparatus
based on a laser displacement sensor, shown in Figure \ref{fig:1}, has been developed.
The thickness is measured along the entire circumference with an angular resolution of 0.1$^{0}$ and
a radial step size of 1 mm, enabling full areal mapping of the target. Each thickness value represents
an average over a circular area with a diameter of 0.5 mm. This method provides a detailed quantification
of thickness variations across the entire target surface.
We report the characterization of thickness variations in FRIB targets with nominal thicknesses ranging from
0.4 to 5 mm.

\section{Apparatus}
The thickness measurement apparatus characterizes the thickness distribution of graphite discs across
a range of nominal thicknesses below 7.4 mm. Detailed specifications for key components
listed in Table \ref{tab:1}. This apparatus
consists of a disc module assembly, a rotational unit, displacement sensors, and a data acquisition (DAQ) system.
The disc module assembly, which includes a ceramic hub (Yttria-stabilized zirconia, YSZ),
graphite disc and spacer, 
is mounted on the rotational unit. An AC motor with a 30:1 reduction gear rotates the graphite disc.
It is driven by an inverter (Hitachi WJ-C1) and controlled by a commercial software (ProDrive Next).
The physical thickness $\textit t$ of the graphite disc is measured using a laser displacement sensor (Keyence CL-3000)
according to the formula \cite{1},
\begin{equation}
  t = C - (A+B)
  \label{eq:1}
\end{equation}
  Here, $C$ is the reference distance between the sensor heads,
while $A$ and $B$ are the measured distances from each sensor to the disc surface. The DAQ system programmed
with LabVIEW synchronously records the disc's thickness and angular position using the displacement sensor
and an incremental encoder (IFM RVP 510), respectively.

\begin{table}
  \centering
  \scriptsize
  \caption{Specifications of Optical unit \cite{2} and Angular encoder \cite{3}.
  }
  \label{tab:1}
  \begin{minipage}[t]{0.65\textwidth}
    \vspace{0pt}
    \centering
    \setlength{\tabcolsep}{6pt}
    \resizebox{\linewidth}{!}{%
      \begin{tabular}{@{}l r@{}}
        \hline
        Optical Unit & CL-L030N \\
        \hline
        Reference distance & 30 mm \\
        Reference measurement range & $\pm 3.7$ mm \\
        Linearity (reference range) & $\pm 0.94\,\mu$m \\
        High precision measurement range & $\pm 1.0$ mm \\
        Linearity (high precision) & $\pm 0.72\,\mu$m \\
        Resolution & $0.25\,\mu$m \\
        Spot diameter & $\varnothing 500\,\mu$m \\
        Laser class & Class 1 \\
        Sampling cycle & 100/200/500/1000 $\mu$s \\
        \hline
        Angular Encoder & RVP510 \\
        \hline
        Resolution & 1--10000 steps \\
        Accuracy & 0.1\textdegree \\
        Electrical design & Incremental \\
        \hline

      \end{tabular}
    }%
  \end{minipage}
  \hfill
 \end{table}

\section{Measurement}

\subsection*{Calibration}\label{Calibration}
In a two-laser thickness measurement system, precise alignment between the target and the lasers is essential
for improving the accuracy of the measurement.
The associated systematic errors are evaluated based on calibration measurements.
One significant source of systematic error in this apparatus is cosine error,
caused by misalignment of the optical axes of the opposing sensor heads.
This error can be minimized by aligning the sensor heads’
optical axes such that they are as collinear and as perpendicular to the target surface as possible \cite{4}.
The calibration procedure is as follows:

1. Set the reference distance C from Equation \ref{eq:1} using a reference
target with known thickness placed between the two sensor heads.

2. Adjust the optical axes to be perpendicular to the closest surface of the reference target.
The measured current from each sensor head will increase as the optical axes approach perpendicular.

3. Adjust optical axes to be collinear. An increase in the currents measured from the four sensor heads,
along with a decrease in the current difference between their readings, can confirm that
the alignment is accurate.

4. Perform a scale calibration using a set of gauge blocks with known thicknesses with accuracy
of $\pm$0.1 ${\mu}m$.The thicknesses of each gauge block are 0.5 mm, 1 mm, 3 mm, and 5 mm.

The measured thicknesses of the gauge blocks using the apparatus show a consistent discrepancy
of $\approx0.0020\ \text{mm}$ from the nominal values. This average discrepancy is programmed
into the DAQ system which automatically subtracts it from all subsequent raw measurements 
Based on the measurement, an offset of 0.01895 mm was applied. The systematic error caused by axis misalignment
after calibration is $\pm 0.0015$ mm. We conservatively set the systematic error at $\pm0.002$ mm,
considering significant figures.

\subsection*{Method}\label{Measurementmethod}
The custom-built thickness measurement apparatus continuously measures the graphite disc’s thickness
while the disc is rotated during the measurement process. The DAQ system records thicknesses and rotational
position data at 0.1$^{0}$ intervals, covering the entire beam-interactive area outer rim (10 mm width)
across nine radial steps of 1 mm. 
The disc rotates at 0.05 Hz (converted from 1.5 Hz motor speed through a 30:1 reduction gear).
This rotation speed was determined as the maximum speed at which the DAQ system can reliably record the data
with minimal signal loss.
The thickness data associated with the rotational position are saved once per revolution. This measurement
is repeated for a total of nine circumferential tracks, starting at the outer edge of the disc and moving inwards
in 1 mm radial steps. Note that each thickness point is the averaged value over a circular area with a 0.5 mm
diameter.  
\begin{figure}
  \begin{center}
    \begin{overpic} [width=0.8\textwidth] {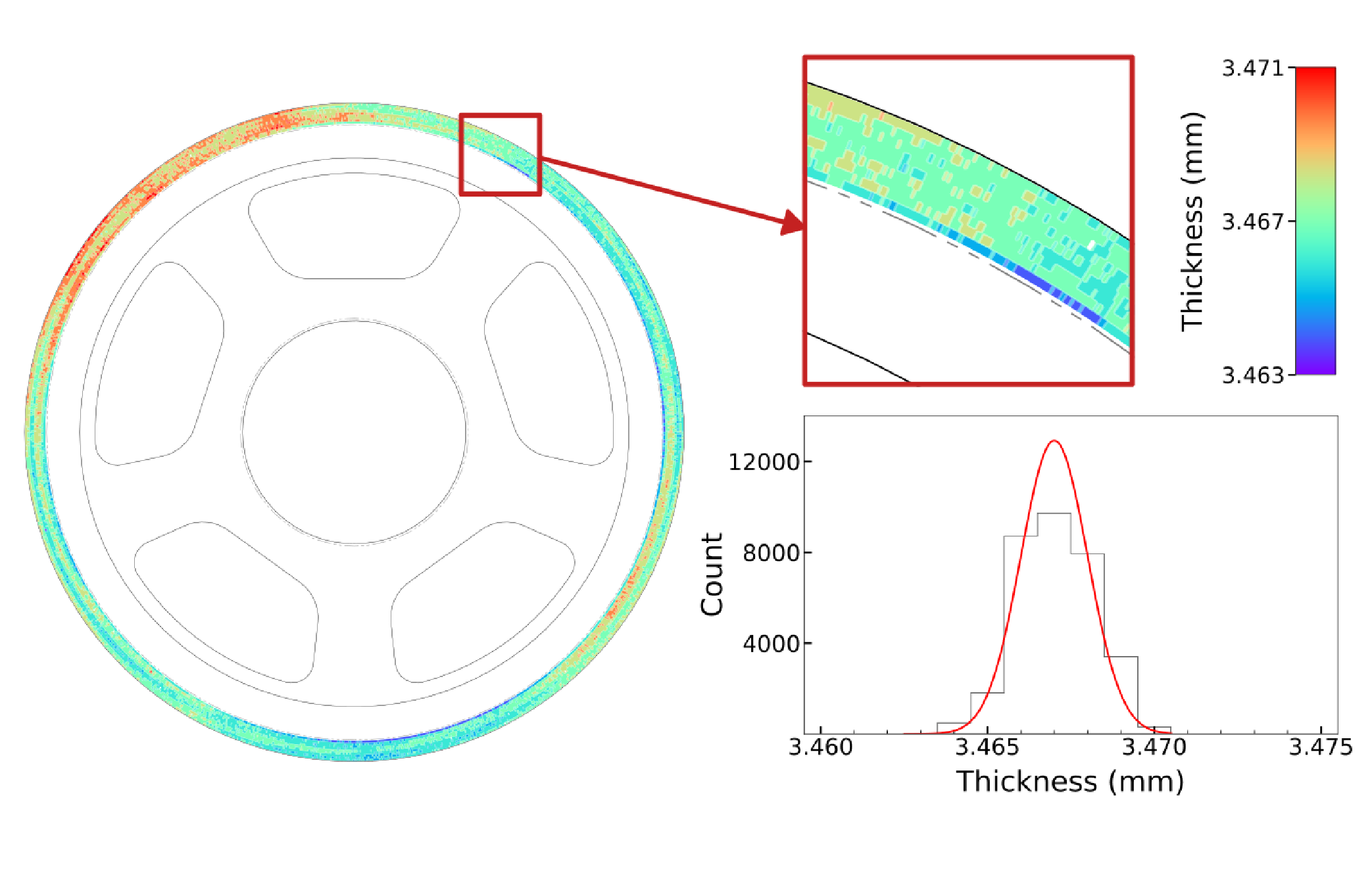}
      \put(0,55.5){\colorbox{white}{\small a)}}
      \put(77.5,55.5){\colorbox{white}{\small b)}}
      \put(59.5,29.5){\colorbox{white}{\small c)}}
    \end{overpic}
    \caption{Representation of 3.5mm graphite disc. a) Two-dimensional contour plot of the
      measured thickness variation across the disc's outer rim (9 mm width) projected onto the graphite target
      diagram. b) Magnified view showing detailed the measurement traces in the selected region of the contour plot.
      c) Histogram
      illustrating the distribution of physical thickness measurements, where the x-axis represents thickness
      and the y-axis indicates the count corresponding to each thickness value measured across the disc.
    }
    \label{fig:2}
  \end{center}
\end{figure}
As an example, the measured thickness distribution of the 3.5 mm disc is presented as two-dimensional contour
map projected onto the disc diagram in Figure \ref{fig:2} a) and a magnified view with a scaled color bar
indicating thickness shown in Figure \ref{fig:2} b). The corresponding thicknesses are plotted as a one-dimensional
histogram to obtain the average thickness and variance shown in Figure \ref{fig:2} c). The histogram shows
the number of thickness measurements
as a function of thickness, and the red line represents the Gaussian fit to the histogram.
The statistical error for each disc is calculated by the Gaussian fit to the histogram of the measured
thickness. For the 3.5 mm disc shown in Figure \ref{fig:2}, the statistical uncertainty ($1\sigma$) is $\pm$0.001 mm.
The total uncertainty including the systematical error is $\pm0.002$ mm, and is given by,   

\begin{equation}
  \Delta t = \sqrt{\Delta t_{st} ^2+ \Delta t_{sys} ^2}
  \label{eq:2}
\end{equation}

As a result, the measured thickness of the 3.5 mm disc shown in Figure \ref{fig:2} is $3.467 \pm0.002$ mm.

\section{Results}\label{Results}

\begin{figure}
  \begin{center}
    \begin{overpic} [width=1\textwidth] {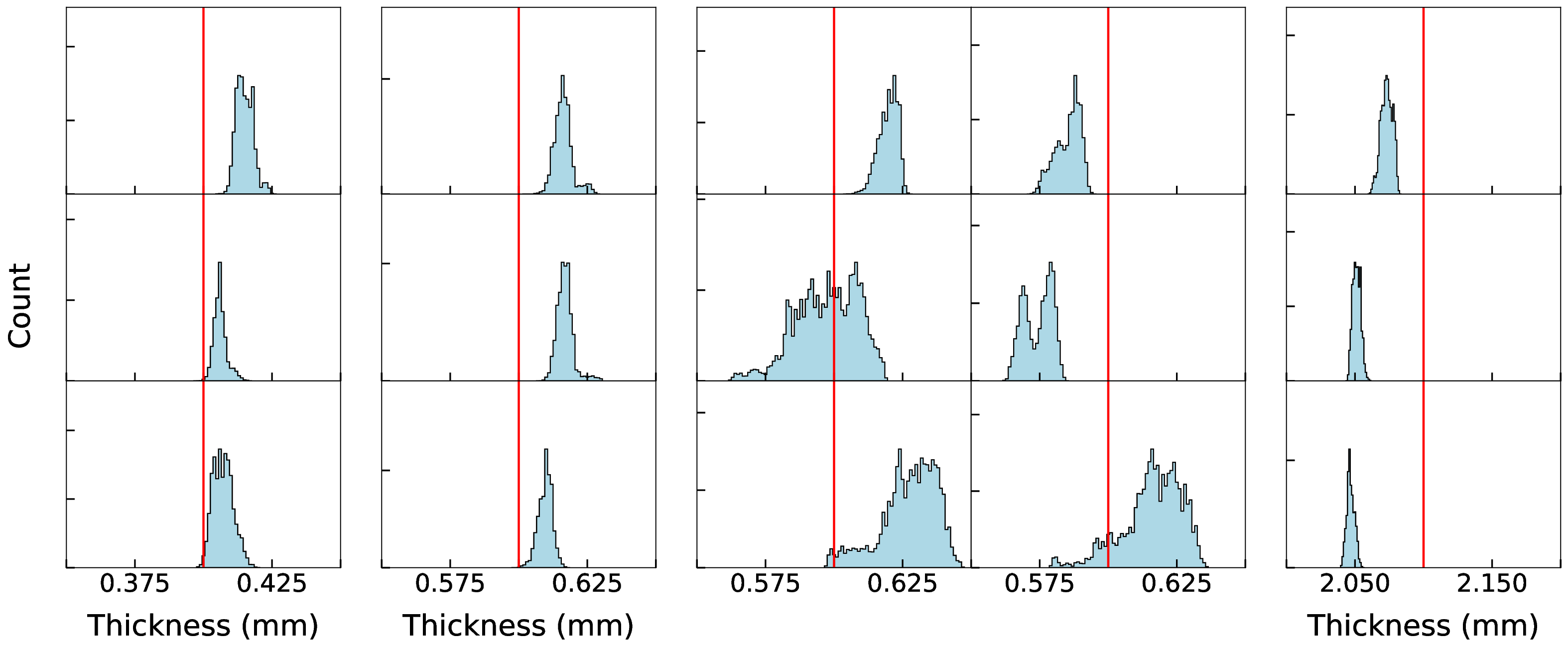}
      \put(8,38){\small a)}
      \put(14.5,38){\color{red}{\small 0.4 mm}}
      \put(27.5,38){\small b)}
      \put(34,38){\color{red}{\small 0.6 mm}}
      \put(47.5,38){\small c)}
      \put(54,38){\color{red}{\small 0.6 mm}}
      \put(85,38){\small d)}
      \put(91.5,38){\color{red}{\small 2.1 mm}}
    \end{overpic}
    \begin{overpic} [width=1\textwidth] {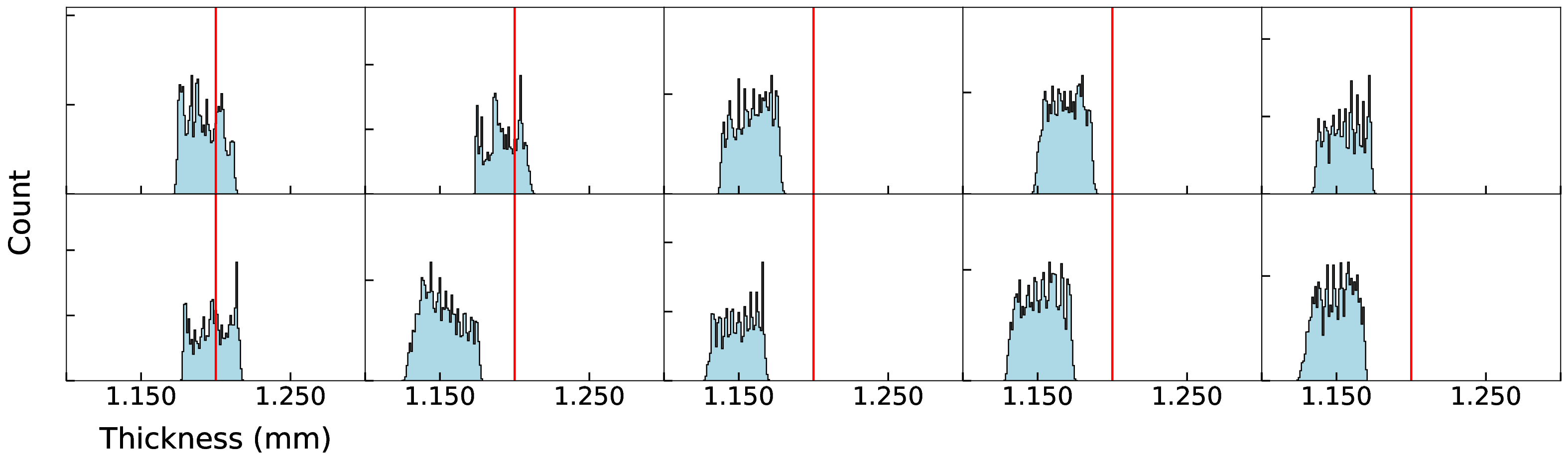}
      \put(8,26){\small e)}
      \put(14.5,26){\color{red}{\small 1.2 mm}}
    \end{overpic}
    \begin{overpic} [width=1\textwidth] {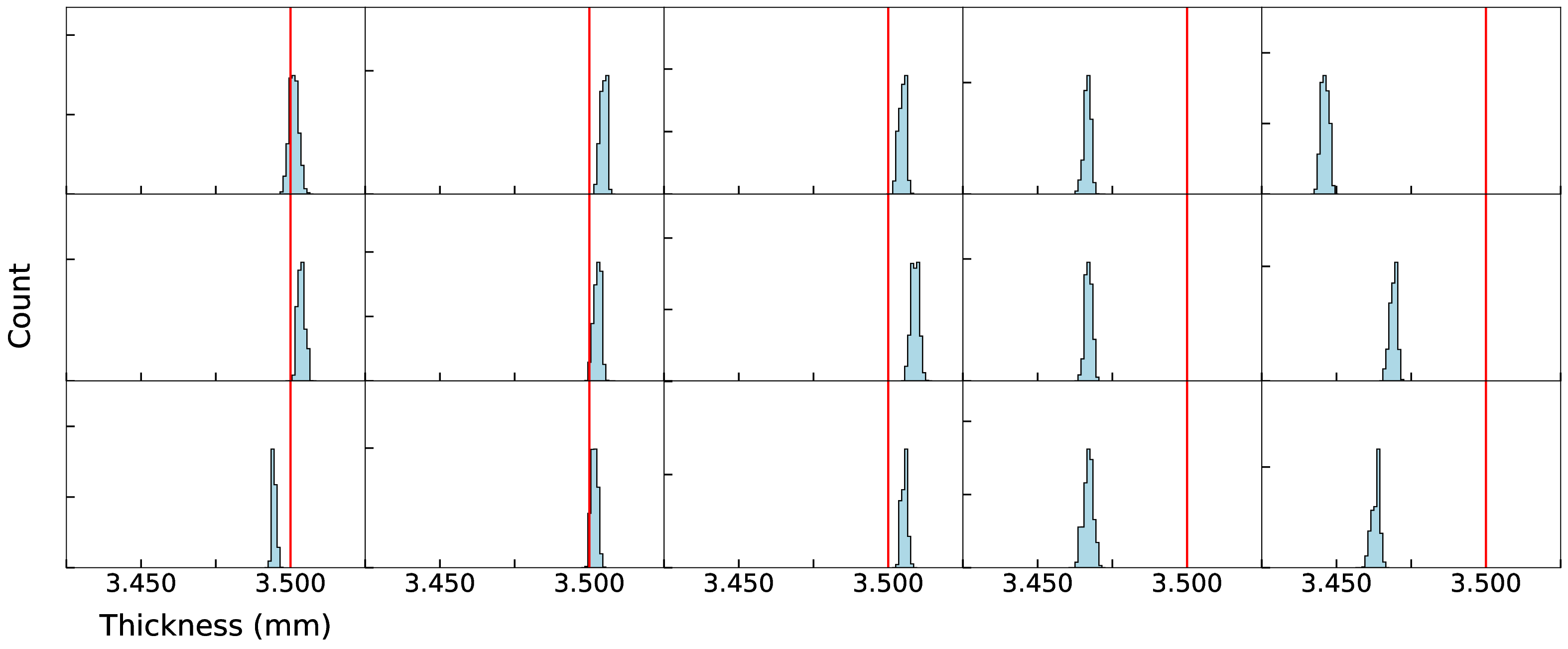}
      \put(8,38){\small f)}
      \put(10.5,38){\color{red}{\small 3.5 mm}}
    \end{overpic}
    \begin{overpic} [width=1\textwidth] {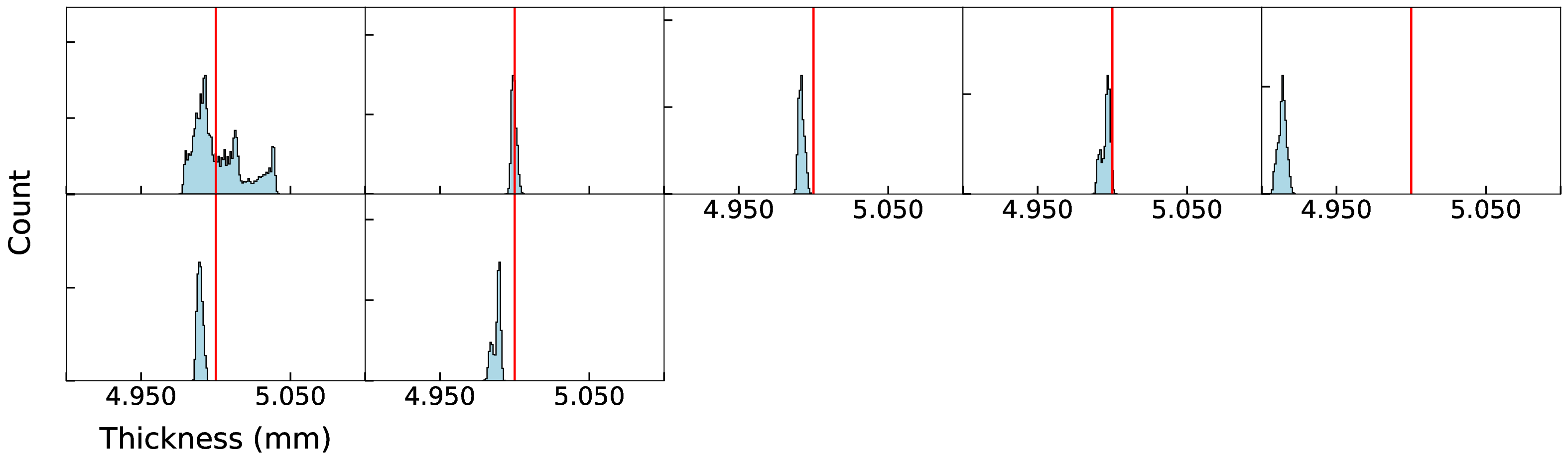}
      \put(8,26){\small g)}
      \put(14.5,26){\color{red}{\small 5.0 mm}}
    \end{overpic}
    \caption{Measured thickness distributions of graphite discs with nominal values of a) 0.4 mm, b-c) 0.6 mm,
      d) 2.1 mm, e) 1.2 mm, f) 3.5 mm and g) 5.0 mm.
      Each panel displays individual measurements across the disc surface, with the x-axis representing the physical
      thickness and the y-axis indicating the count of each thickness value. The red vertical line denotes
      the nominal thickness for each disc.
      Discs a) and b) were manufactured by Supplier A. All Others were manufactured by Supplier B.
    }
    \label{fig:3}
  \end{center}
\end{figure}

The thicknesses of the graphite discs with nominal values of 0.4, 0.6, 1.2, 2.1, 3.5 and 5.0
mm were measured using the custom-built thickness measurement apparatus.
The thickness distributions measured for all prototypes and spare production targets are shown in Figure \ref{fig:3}.
Discs with different nominal thicknesses are grouped separately, and the nominal thickness for each group is
indicated by the red line in the corresponding panel.
The thin discs (0.4 mm and 0.6 mm) are prototypes for future multi-slice production target.
Thicker discs are currently used in a single-slice production configurations.

The 0.4 mm discs are shown in Figure \ref{fig:3} a), with the x-axis ranging from 0.35 mm to 0.45 mm.
All samples were manufactured by Supplier A.
The average thickness of the three 0.4 mm graphite discs is $\approx$ 0.41 mm, with low variance
(1$\sigma$ $\approx 0.004$ mm). The thicknesses were intentionally manufactured
slightly above the nominal value, as machining was stopped once the nominal thickness was achieved.
The thickness variance of 0.005 mm is considered acceptable for operational use.
The 0.6 mm discs manufactured by Supplier A and Supplier B are presented in Figure \ref{fig:3} b) and
Figure \ref{fig:3} c), respectively, with the x-axis spanning 0.55 to 0.65 mm.
The thicknesses of supplier A samples are slightly above the nominal value with low variance
(1$\sigma$ $\approx 0.004$ mm), consistent with the trend observed for the 0.4 mm discs.
In contrast, the thicknesses of the supplier B samples show larger fluctuations from the nominal values,
with relatively high variance (1$\sigma$ from 0.004 to 0.011 mm).
While some samples are within the acceptable range for operational use,
others exceed the tolerable variation threshold.
The 1.2 mm disc manufactured by supplier B are illustrated in Figure \ref{fig:3} e) with x-axis ranging
from 1.1 mm to 1.3 mm.
Similar to 0.6 mm discs, the thickness varied between individual discs.
Although some discs are slightly thicker than
nominal value, most discs are thinner.
The variance (1$\sigma$ $>$ 0.01 mm) exceeds the acceptable threshold for all discs.
This deviation is attributed to abnormal machining, which results in a consistent trend of decreasing thickness
from the outer edge toward the center (see Section \ref{cd}).
The 2.1, 3.5 and 5 mm discs manufactured by supplier B, are shown in the panels of d), f) and g)
of Figure \ref{fig:3}, respectively.
The thicknesses consistently fluctuate; however the variances remain low (1$\sigma \leq 0.004$)
except for the first disc of the 5.0 mm thick discs.
Notably, the  3.5 mm discs demonstrate the highest thickness precision among all discs with a
1$\sigma$ $\approx$ 0.002 mm.
The average thickness measurements and associated uncertainties for each disc
are summarized in Table \ref{tab:1}. Discs with
same nominal thickness and manufacturer are grouped together.
Each entry includes the discs' serial number (SN), average thickness, and corresponding uncertainty, which is
calculated according to Equation \ref{eq:2}.
The missing serial numbers indicate that the corresponding discs were used in operation.

\begin{table}
  \centering
  \scriptsize
  \caption{
    Measured average thicknesses and associated uncertainties for FRIB graphite discs.
    SN and unc denote serial number and measurement uncertainty, respectively.
  }
  \label{tab:2}
  \begin{minipage}[t]{0.75\textwidth}
    \vspace{0pt}
    \centering
    \setlength{\tabcolsep}{4pt}
    \resizebox{\linewidth}{!}{%
      \begin{tabular}{@{}rrr |  rrr@{}}
        \hline
        \multicolumn{3}{c|}{0.4 mm (Supplier A)} &\multicolumn{3}{c}{0.6 mm (Supplier A)} \\
	\hline
        SN & Average thickness (mm) & unc. (mm) & SN & Average thickness (mm) & unc. (mm) \\
        \hline
        1 & 0.415 & 0.004 & 1 & 0.616 & 0.004 \\
        2 & 0.406 & 0.003 & 2 & 0.617 & 0.004 \\
        3 & 0.407 & 0.004 & 3 & 0.610 & 0.004 \\
        \hline
        \hline
        \multicolumn{3}{c|}{0.6 mm (Supplier B)} & \multicolumn{3}{c}{5.0 mm (Supplier B)} \\
        \hline
	SN & Average thickness (mm) & unc. (mm) & SN & Average thickness (mm) & unc. (mm) \\
	\hline
        1 & 0.620 & 0.004 & 4  & 5.002 & 0.017 \\
        2 & 0.597 & 0.011 & 5  & 4.989 & 0.003 \\
        3 & 0.627 & 0.010 & 6  & 5.000 & 0.003 \\
        4 & 0.585 & 0.005 & 7  & 4.988 & 0.004 \\
        5 & 0.574 & 0.005 & 8  & 4.992 & 0.003 \\
        6 & 0.616 & 0.011 & 9  & 4.996 & 0.004 \\
          &       &       & 10 & 4.914 & 0.004 \\
        \hline
 	\multicolumn{3}{c|}{1.2 mm (Supplier B)} & \multicolumn{3}{c}{3.5 mm (Supplier B)} \\
	\hline
      SN & Average thickness (mm) & unc. (mm) & SN & Average thickness (mm) & unc. (mm) \\
      \hline
       2 & 1.192 & 0.011 &  2 & 3.501 & 0.003 \\
       3 & 1.198 & 0.011 &  3 & 3.504 & 0.002 \\
       4 & 1.192 & 0.010 &  4 & 3.495 & 0.002 \\
       5 & 1.151 & 0.013 &  5 & 3.505 & 0.002 \\
       6 & 1.159 & 0.011 &  6 & 3.503 & 0.002 \\
       7 & 1.150 & 0.011 &  7 & 3.502 & 0.002 \\
       8 & 1.169 & 0.010 &  8 & 3.505 & 0.002 \\
       9 & 1.153 & 0.012 &  9 & 3.509 & 0.002 \\
      10 & 1.156 & 0.011 & 10 & 3.505 & 0.002 \\
      11 & 1.150 & 0.011 & 11 & 3.467 & 0.002 \\
      \cline{1-3}
      \multicolumn{3}{c|}{2.1 mm (Supplier B)} & 12 & 3.467 & 0.002 \\
      \cline{1-3}
       SN & Average thickness (mm) & unc. (mm) & 13 & 3.467 & 0.003\\
       \cline{1-3}
       3 & 2.073 & 0.004 & 14 & 3.446 & 0.002 \\
       4 & 2.051 & 0.004 & 15 & 3.469 & 0.002 \\
       5 & 2.047 & 0.004 & 16 & 3.463 & 0.002 \\
      \hline
      \end{tabular}
    }%
  \end{minipage}
  \hfill
\end{table}

\begin{figure}
  \begin{center}
    \begin{overpic} [width=0.49\textwidth] {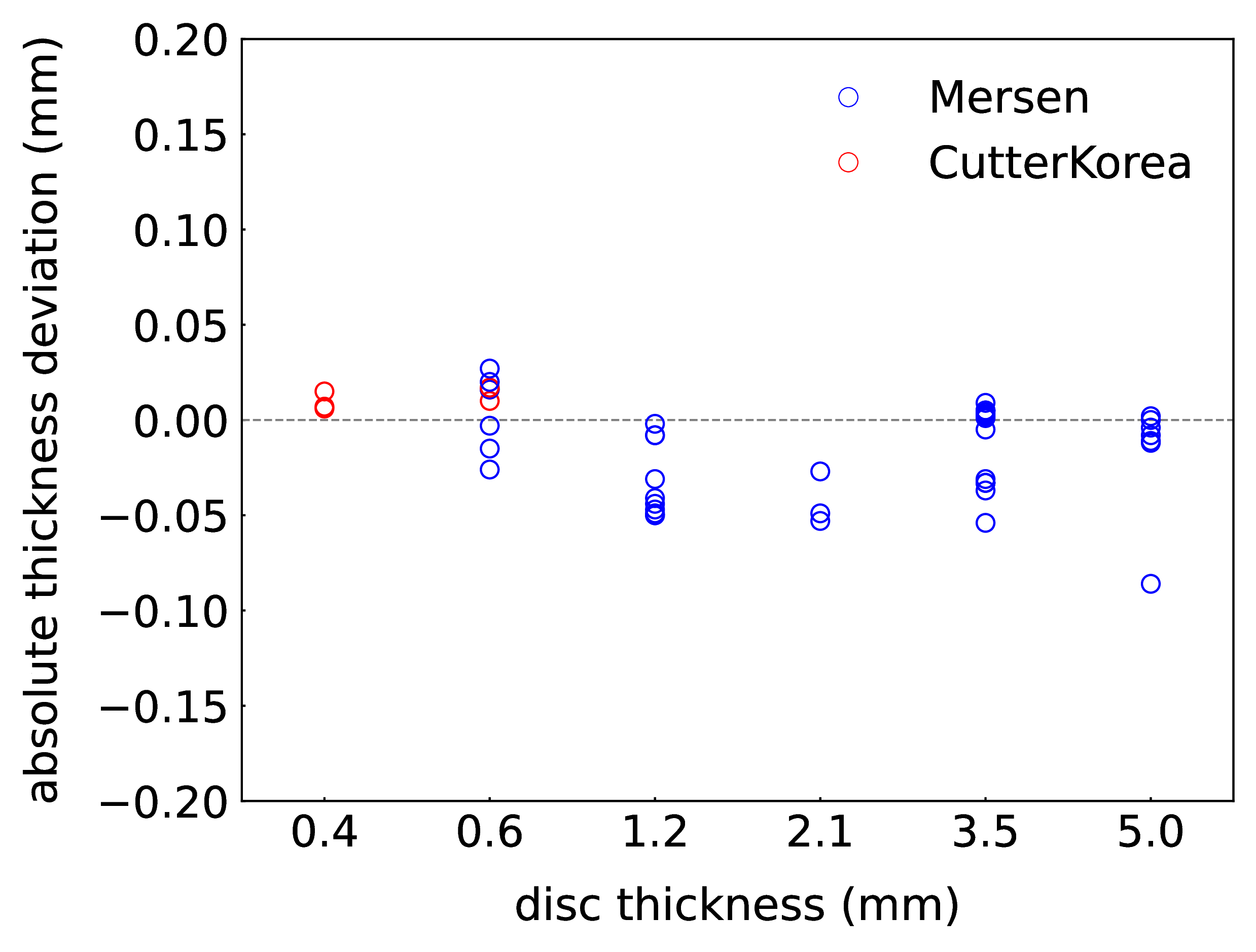}
      \put(20,67){\colorbox{white}{\small a)}}
      \put(66,67){\colorbox{white}{\textcolor{red}{$\circ$}\scriptsize ~~ Supplier A}}
      \put(66,61){\colorbox{white}{\textcolor{blue}{$\circ$}\scriptsize ~~ Supplier B ~ ~ }}
    \end{overpic}
    \begin{overpic} [width=0.49\textwidth] {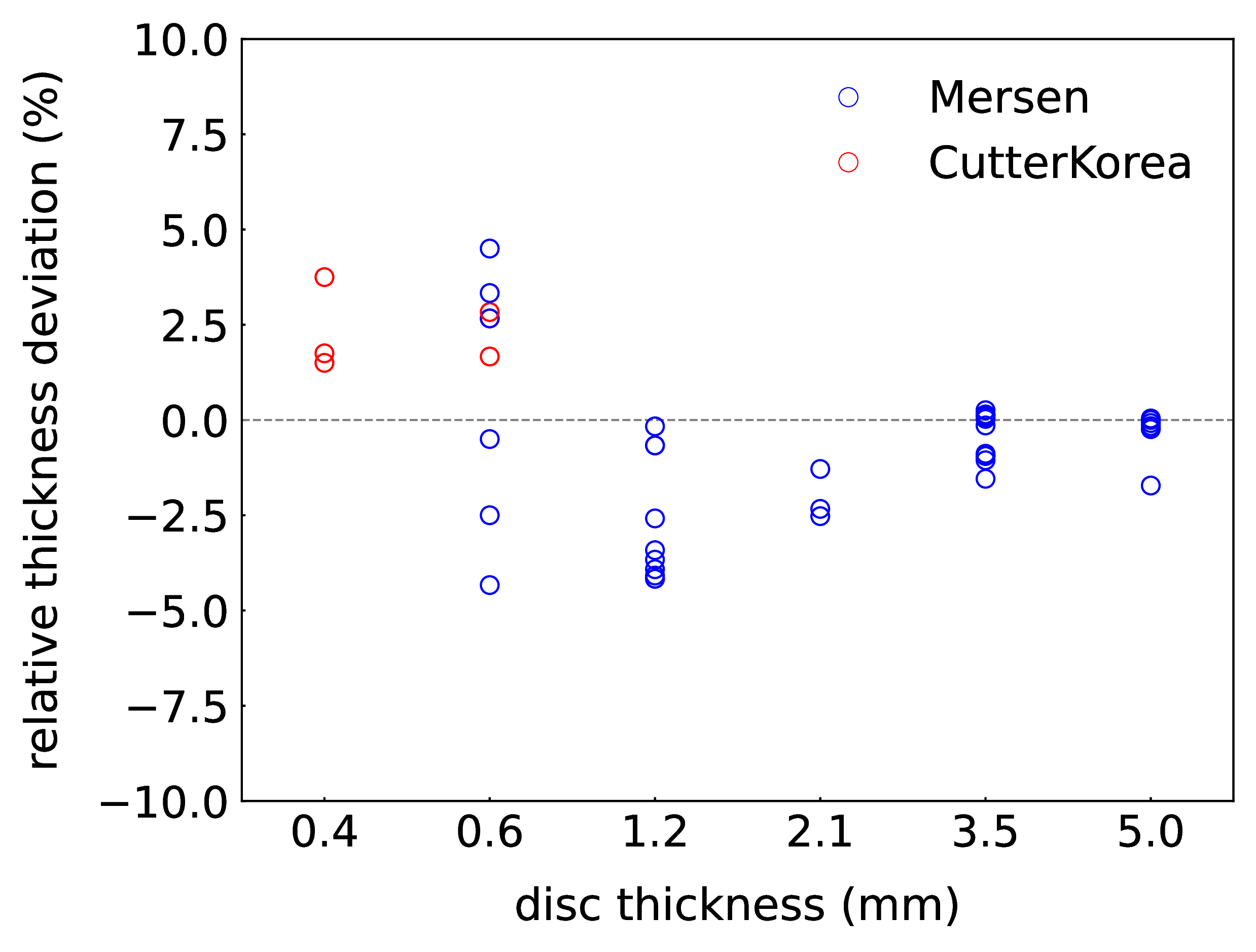}
      \put(20,67){\colorbox{white}{\small b)}}
      \put(66,67){\colorbox{white}{\textcolor{red}{$\circ$}\scriptsize ~~ Supplier A}}
      \put(66,61){\colorbox{white}{\textcolor{blue}{$\circ$}\scriptsize ~~ Supplier B ~ ~ }}
    \end{overpic}
    \begin{overpic} [width=0.498\textwidth] {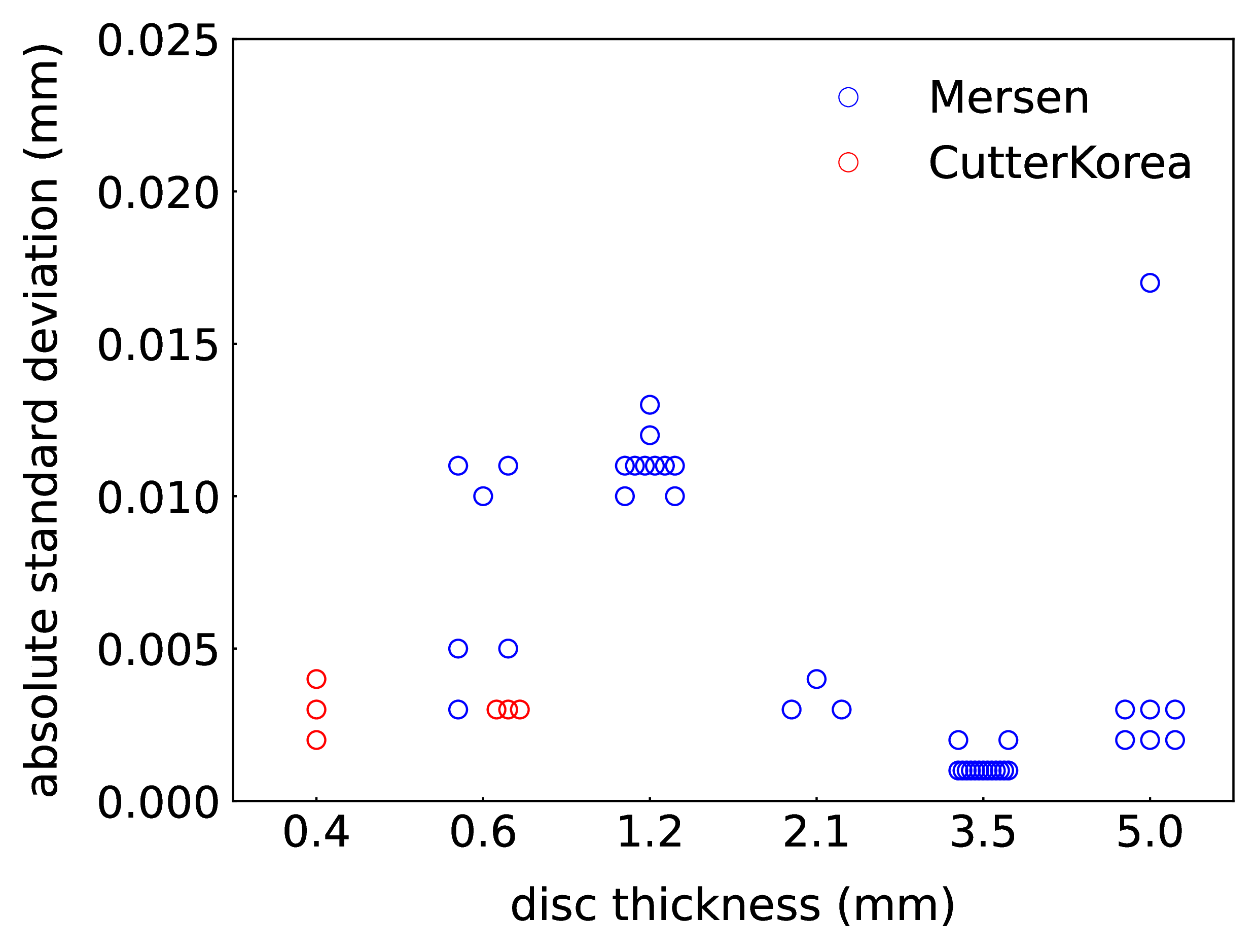}
      \put(20,67){\colorbox{white}{\small c)}}
      \put(66,67){\colorbox{white}{\textcolor{red}{$\circ$}\scriptsize ~~ Supplier A}}
      \put(66,61){\colorbox{white}{\textcolor{blue}{$\circ$}\scriptsize ~~ Supplier B ~ ~ }}
    \end{overpic}
    \begin{overpic} [width=0.49\textwidth] {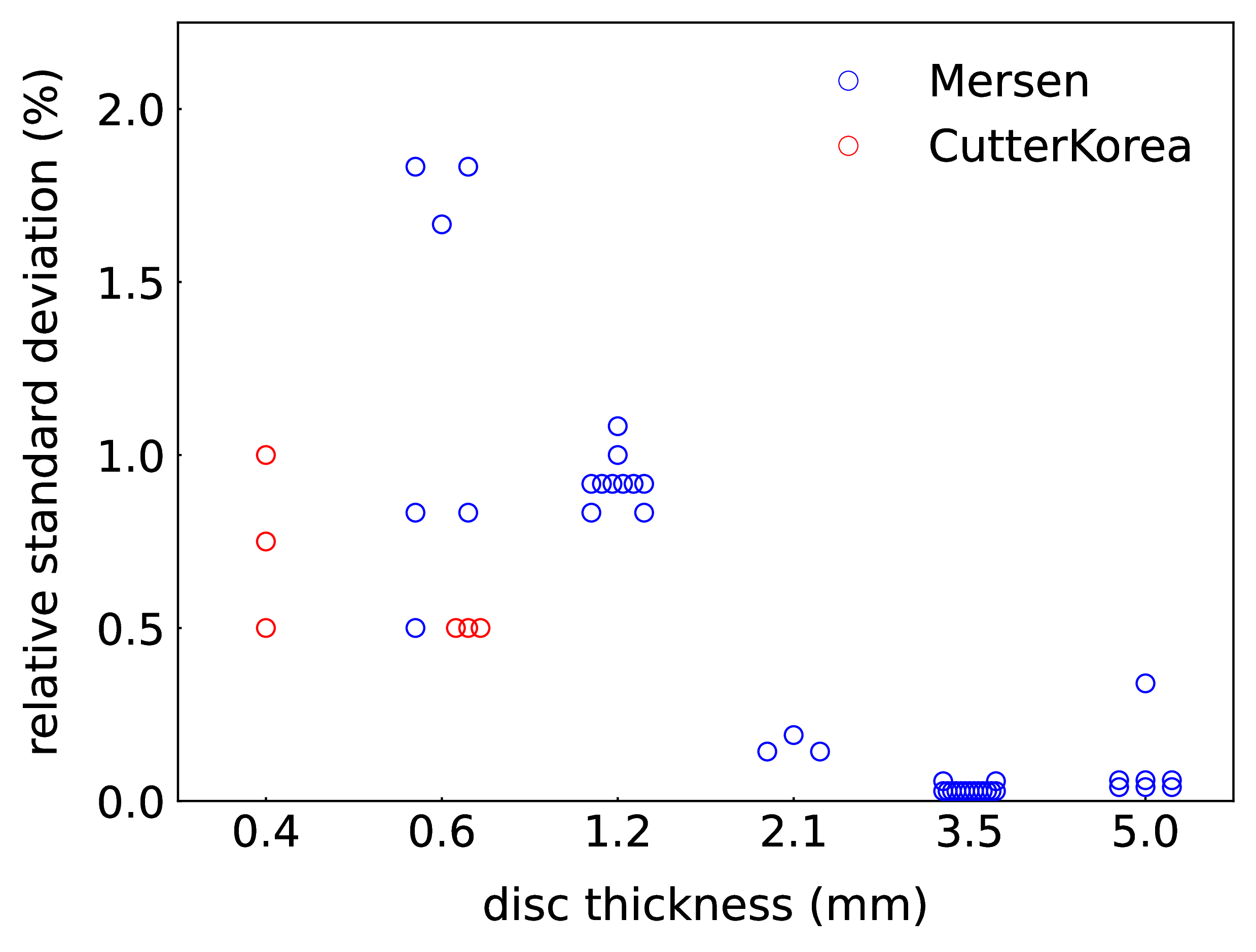}
      \put(16,67){\colorbox{white}{\small d)}}
      \put(66,67){\colorbox{white}{\textcolor{red}{$\circ$}\scriptsize ~~ Supplier A}}
      \put(66,61){\colorbox{white}{\textcolor{blue}{$\circ$}\scriptsize ~~ Supplier B ~ ~ }}
    \end{overpic}
    \caption{
     Graphite discs manufactured by supplier A and B are represented by red and blue circles, respectively.
     a) Absolute and b) relative deviations of the average measured thickness from the nominal value, expressed
     in mm and \%.
     c) Standard deviation of the measured thickness for each disc, expressed in mm and d) in \%.
    }
	\label{fig:4}
  \end{center}
\end{figure}

A comparative analysis of the target thickness accuracy for disc manufactured by supplier A and B
represented by red and blue circles respectively, is illustrated
in Figure \ref{fig:4}.
Figure \ref{fig:4} a) and b) present the absolute and relative deviations of the average measured thickness
from the nominal values, expressed in mm and $\%$.
Supplier B discs consistently show negative deviations in most cases, indicating a systematic tendency to be thinner than
specified. In contrast, supplier A discs exhibit slightly positive or near-zero deviations,
indicating closer agreement with nominal thicknesses.
In terms of relative deviation, supplier B discs with nominal thicknesses below 2.1 mm display larger discrepancies
than thicker ones. Supplier A discs, on the other hand, show relatively uniform deviations across all thicknesses
between the 0.4 mm and 0.6 mm cases.
Figure \ref{fig:4} c) and d) show the standard deviation of the measured thickness for each disc,
expressed in mm and $\%$.

For both supplier A and supplier B discs, there is no clear correlation between disc thickness and the absolute
standard deviation in mm, except at 1.2 mm (see Appendix).
This indicates that the absolute standard deviation is primarily constrained by the machining tolerance,
which is determined by the manufacturer's production capabilities.
The typical machining tolerance is approximately 5 $\mu$m.
As a result, the relative standard deviation becomes inversely proportional to the disc thickness.
This observation highlights inherent limitations in the current manufacturing process.
A thickness of 0.4 mm appears to define the practical lower limit for achieving a relative standard deviation
(2$\sigma$) below 2 $\%$.

\section{Conclusion}
\label{cd}
We present the thickness characterization of FRIB production targets with nominal thicknesses of 0.4, 0.6,
1.2, 2.1, 3.5, and 5 mm using a custom-built non-contact thickness measurement apparatus.
This study evaluates both thickness accuracy and variation. The thickness accuracy with respect to nominal
values is $\approx$ $\pm$ (1-5)$\%$, with absolute variations ranging from 1 to 10 $\mu$m.
No strong correlation is observed between thickness variation and nominal thickness, indicating that
mechanical tolerances are consistently maintained across the full range of nominal thicknesses.
However, the thickness variation is inversely proportional to the nominal thickness of the targets,
since a fixed absolute deviation results in a higher percentage error at thinner nominal thickness.
Most spare discs in the single-slice target range of 1.2 to 5 mm satisfy the relative standard deviation criterion
(2$\sigma$) within 2$\%$.
A 0.4 mm thick disc represents the practical lower limit achievable under current mechanical tolerances
, while maintaining a required standard deviation (2$\sigma$) within 2$\%$.
This information will support future target design
decisions for both single-slice and multi-slice graphite discs and provide a baseline for evaluating
fabrication precision.

\appendix

\section*{Appendix}
\label{appendix:1}
\begin{figure}[h!]
  \begin{center}
    \begin{overpic} [width=0.8\textwidth] {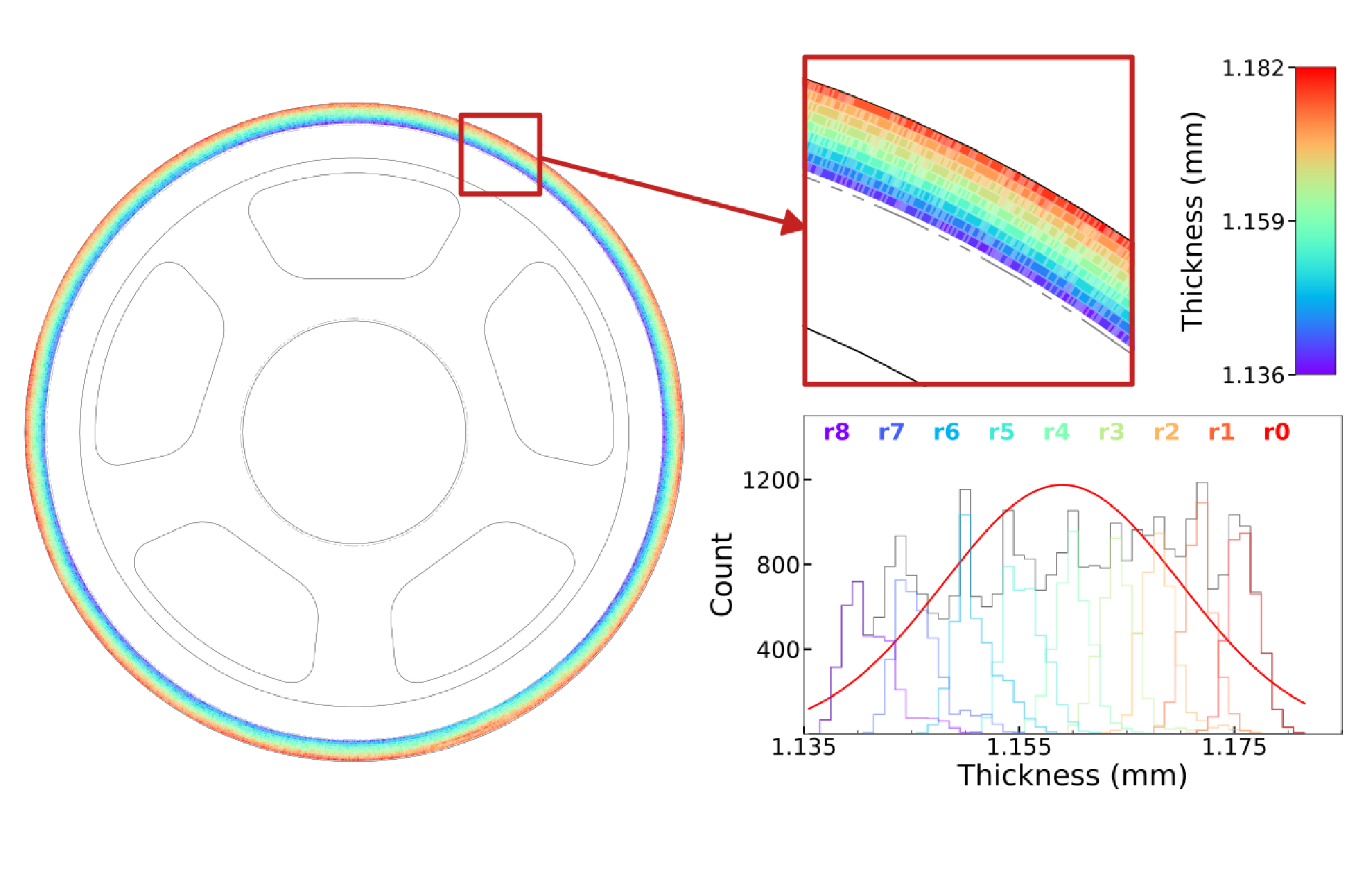}
      \put(0,55.5){\colorbox{white}{\small a)}}
      \put(77.5,55.5){\colorbox{white}{\small b)}}
      \put(59.5,27.5){\colorbox{white}{\small c)}} 
    \end{overpic}
    \caption{
      Thickness profile analysis of a 1.2 mm graphite disc indicating radial variations.
      a) Two-dimensional contour map of the measured thickness across the disc's outer rim (9 mm width)
      , projected onto the graphite target schematic.
      b) Magnified view showing detailed measurement traces on the contour plot.
      c) The measured thickness distributions across the disc. r0 to r8 represent discrete radial positions
      from the outer edge toward the center (1 mm step).
    }
    \label{fig:5}
  \end{center} 
\end{figure}

The standard deviations observed for the 1.2 mm thick graphite discs are significantly higher than those of all other
thicknesses. As illustrated in Figure \ref{fig:5}, a detailed analysis of one representative disc shows a gradual decrease
in thickness from the outer edge toward the center. The distributions presented in Figure \ref{fig:5} c)
are distinctly separated across the radial positions r0 to r8, indicating a systematic radial variation
rather than random fluctuation.
This indicates that the mechanical machining process resulted in a radial gradient in thickness,
which was not expected under nominal manufacturing conditions.
However, since the primary beam size at the target  is 0.25 mm (1$\sigma$), these discs can still be used for operations.
This is supported by the fact that the relative standard deviation within a 3 mm-wide region remains approximately 1$\%$.

\section*{ACKNOWLEDGMENTS}
This material is based upon work supported by the U.S. Department of Energy, Office of Science, Office of Nuclear
Physics and used resources of the Facility for Rare Isotope Beams (FRIB) Operations, which is a DOE Office
of Science User Facility under Award Number DE-SC0023633.



\begin{thebibliography}{99}
\bibitem{int_1}
  J. Wei \textit{et al.}, "Technological developments and accelerator improvements
  for the FRIB beam power ramp-up"
  , 2024 \textit{JINST} 19 T05011,
  \url{https://dx.doi.org/10.1088/1748-0221/19/05/T05011}
  
\bibitem{int_2}
  P. Ostroumov \textit{et al.}, "Acceleration of uranium beam to record power of 10.4 kW and
  observation of new isotopes
  at Facility for Rare Isotope Beams"
  , \textit{Phys. Rev. Accel. Beams}, 27, 089901 (2024),
  \url{https://link.aps.org/doi/10.1103/PhysRevAccelBeams.27.060101}
  
\bibitem{int_3_1}
  F. Pellemoine \textit{et al.}, "Thermo-mechanical behaviour of a single slice test device for the FRIB high power target"
  , \textit{Nucl. Instrum. Methods Phys. Res. A}, 655, 3-9 (2011),
  \url{https://www.sciencedirect.com/science/article/pii/S0168900211011211}

\bibitem{int_3_2}
  F. Pellemoine \textit{et al.}, "Development of a production target for FRIB: thermo-mechanical studies"
  , \textit{Journal of Radioanalytical and Nuclear Chemistry}, 299, 933-939 (2014),
  \url{https://doi.org/10.1007/s10967-013-2623-7}

  \bibitem{int_3_3}
  J. Song \textit{et al.}, "A SINGLE-SLICE ROTATING GRAPHITE TARGET AT FRIB"
  , \textit{Proc. HIAT25}, East Lansing, MI, June 2025,
  \url{https://meow.elettra.eu/82/pdf/TUB01.pdf}

    \bibitem{int_4}
  M. Avilov \textit{et al.}, "A 50-kW prototype of the high-power production target for the FRIB"
  , \textit{Journal of Radioanalytical and Nuclear Chemistry}, 305, 817-823, 2015,
  \url{https://doi.org/10.1007/s10967-014-3908-1}

  \bibitem{int_5_1}
  N. Simons \textit{et al.}, "Proton irradiated graphite grades for a long baseline neutrino facility experiment"
  , \textit{Phys. Rev. Accel. Beams},
  20, 071002, 2017,
  \url{https://link.aps.org/doi/10.1103/PhysRevAccelBeams.20.071002}

  \bibitem{int_5_2}
    N. Simons \textit{et al.}, "120 GeV neutrino physics graphite target damage assessment using electron microscopy and high-energy x-ray diffraction"
    , \textit{Phys. Rev. Accel. Beams}
    , 22, 041001, 2019,
    \url{https://link.aps.org/doi/10.1103/PhysRevAccelBeams.22.041001}
    
      \bibitem{int_5_3}
  S. Makimura \textit{et al.}, "Present status of construction for the muon target in J-PARC"
  , \textit{Nuclear Instruments and Methods in Physics Research A}
  , 600, 146-149, 2009,
  \url{https://www.sciencedirect.com/science/article/pii/S0168900208016859}
  
 \bibitem{int_5_4}
  S. Makimura \textit{et al.}, "Development of a Muon Rotating Target for J-PARC/MUSE"
  , \textit{Physics Procedia}
  , 32, 795-801, 2012,
  \url{https://www.sciencedirect.com/science/article/pii/S1875389212010553}

  \bibitem{int_5_5}
  U. Koster \textit{et al.}, "Off-line production of intense $^{7,10}$Be$^{+}$ beams"
  , \textit{Nuclear Instruments and Methods in Physics Research B}
  , 204, 343-346, 2003,
  \url{https://www.sciencedirect.com/science/article/pii/S0168583X02019523}

 \bibitem{int_5_6}
  D. Kiselev \textit{et al.}, "The PSI meson target facility and its upgrade IMPACT-HIMB"
  , \textit{EPJ Web of Conferences}
  , 285, 07002, 2023,
  \url{https://doi.org/10.1051/epjconf/202328507002}

   \bibitem{int_6_1}
  M. Hausmann \textit{et al.}, "Design of the Advanced Rare Isotope Separator ARIS at FRIB"
  , \textit{Nuclear Instruments and Methods in Physics Research Section B}
  , 317, 349-353, 2013,
  \url{https://www.sciencedirect.com/science/article/pii/S0168583X13007210}
  
   \bibitem{int_6_2}
  M. Portillo \textit{et al.}, "Commissioning of the Advanced Rare Isotope Separator ARIS at FRIB"
  , \textit{Nuclear Instruments and Methods in Physics Research Section B}
  , 540, 151-157, 2023,
  \url{https://www.sciencedirect.com/science/article/pii/S0168583X23001556}

  
\bibitem{1}
KEYENCE Corporation,
\textit{How to Measure: Understanding Displacement Sensors / Measurement Systems},
Technical Guide, p.4.
\url{https://www.keyence.com/} 

  
\bibitem{2}
KEYENCE Corporation,
\textit{Confocal Displacement Sensor CL-3000 Series},
Technical Guide, p. 23.
\url{https://www.keyence.com/} 
  
\bibitem{3}
IFM, \emph{RVP510 Data Sheet}, Instrumentation Information and Fabrication Manual (IIFM), 
\url{https://www.ifm.com/us/en/product/RVP510#details}

\bibitem{4}
KEYENCE Corporation,
\textit{User's Manual CL-3000 Series},
Technical Guide, p. 4-16.
\url{https://www.keyence.com/} 


\end{thebibliography}
\end{document}